\documentclass[aip,apl,reprint,twocolumn]{revtex4-1}

\usepackage{graphicx}
\usepackage{amsmath}
\usepackage{amssymb}
\usepackage[colorlinks=true,citecolor=blue,linkcolor=blue]{hyperref}
\usepackage{amsfonts}
\usepackage[latin9]{inputenc}
\usepackage{color}
\usepackage{textcomp}
\usepackage{multirow}
\usepackage{xcolor}
\usepackage{lmodern}
\usepackage[T1]{fontenc}
\usepackage{makecell}

\renewcommand{\vec}[1]{\ensuremath{\boldsymbol{#1}}}
\begin{document}

\title{Graphene quantum blisters: a tunable system to confine charge carriers}
\date{\today }
\author{H. M. Abdullah}
\email{alshehab211@gmail.com}
\affiliation{Department of Physics, King Fahd University of Petroleum and Minerals, 31261 Dhahran, Saudi Arabia}
\affiliation{Saudi Center for Theoretical Physics, P.O. Box 32741, Jeddah 21438, Saudi Arabia}
\affiliation{Department of Physics, University of Antwerp, Groenenborgerlaan 171, B-2020 Antwerp, Belgium}
\author{M. Van der Donck}
\affiliation{Department of Physics, University of Antwerp, Groenenborgerlaan 171, B-2020 Antwerp, Belgium}
\author{H. Bahlouli}
\affiliation{Department of Physics, King Fahd University of Petroleum and Minerals, 31261 Dhahran, Saudi Arabia}
\affiliation{Saudi Center for Theoretical Physics, P.O. Box 32741, Jeddah 21438, Saudi Arabia}
\author{F. M. Peeters}
\affiliation{Department of Physics, University of Antwerp, Groenenborgerlaan 171, B-2020 Antwerp, Belgium}

\author{B. Van Duppen}
\email{ben.vanduppen@uantwerpen.be}
\affiliation{Department of Physics, University of Antwerp, Groenenborgerlaan 171, B-2020 Antwerp, Belgium}

\begin{abstract}
Due to Klein tunneling, electrostatic confinement of electrons in graphene is not possible. This hinders the use of graphene for quantum dot applications. Only through quasi-bound states with finite lifetime has one achieved to confine charge carriers. Here we  propose that bilayer graphene with a local region of decoupled graphene layers is able to generate bound states under the application of an electrostatic gate. The discrete energy levels in such a quantum blister  correspond to localized electron and hole states in the top and bottom layers. We find that this layer localization and  the energy spectrum itself are tunable by a global electrostatic gate and that the latter also coincides with the electronic  modes in a graphene disk. Curiously, states with energy close to the continuum  exist primarily in the classically forbidden region outside the domain defining the blister. The results are robust against variations in size and shape of the blister which shows that it  is a versatile system to achieve tunable electrostatic confinement in graphene. 
\end{abstract}

\maketitle

Ever since the discovery of graphene, researchers have tried to confine electrons in graphene-based quantum dots (QDs)\cite{Guettinger2012} because of the vast range of new applications for QDs in for instance electronic circuitry\cite{Gueclue2013}, photovoltaics\cite{Bacon2013}, qubits\cite{Trauzettel2007}, and gas sensing\cite{Sun2013}. Graphene as a basis for these QDs could enable fast and flexible devices. On a more fundamental level, the ultra-relativistic nature of graphene charge carriers made researchers wonder how they would respond to confinement\cite{Rozhkov_2011}. It is, however, exactly this peculiar property that prohibits the use of traditional QD fabrication techniques such as local electrostatic gating to confine carriers. The Klein tunnelling effect\cite{Katsnelson2006} allows electrons to use hole states in the gated region to escape the QD. The graphene quantum blister (GQB), proposed in this Article, overcomes this limitation and acts as a tunable graphene quantum dot that still harnesses the peculiar electronic properties of graphene.

The quest to confine Dirac Fermions in graphene QDs has resulted in many propositions. For instance, one has tried using magnetic fields\cite{Espinosa-Ortega2013,Martino2007}, cutting the flake into small nanostructures\cite{Mirzakhani2016,Zebrowski2013} or using the substrate to induce a band gap\cite{Gutierrez-Rubio2015,Recher2009}. However, magnetic fields bring along many difficulties in nano-sized systems \cite{Liu2017}, QDs made from nanostructures are highly sensitive to the precise shape of the edge, which is hard to control \cite{Espinosa-Ortega2013}, and also the band gap produced in graphene by a substrate is very difficult to control\cite{Decker2011}. Due to these difficulties experimental realization of graphene QDs is limited and this hinders applicability.

This has, however, not withheld researchers from trying to apply extreme external conditions. Under high magnetic field\cite{Jung2011}  or supercritical charges\cite{Mao2016} confinement was realized, but only  quasi-bound states with a relatively short life time\cite{Matulis2008} were observed. Recently, a few experiments\cite{Zhao2015,Ghahari2017,Gutierrez2016,Lee2016,Freitag2016} were conducted to detect short-lived quasi-bound states in single layer graphene   by using advanced  substrate engineering and the incorporation of an electrostatic potential induced by the tip of the scanning tunneling electron microscope (STM).  There is only one recent experiment\cite{Qiao2017} that realized bound states with a longer lifetime in  a QD in a graphene sheet through a strong coupling between the graphene sheet and the substrate. However, the bound states are only tunable through careful controlling of the distance between the sample and the STM tip.

\begin{figure}[t!]
\centering\graphicspath{{./Figures/}}
\includegraphics[width=2.8 in]{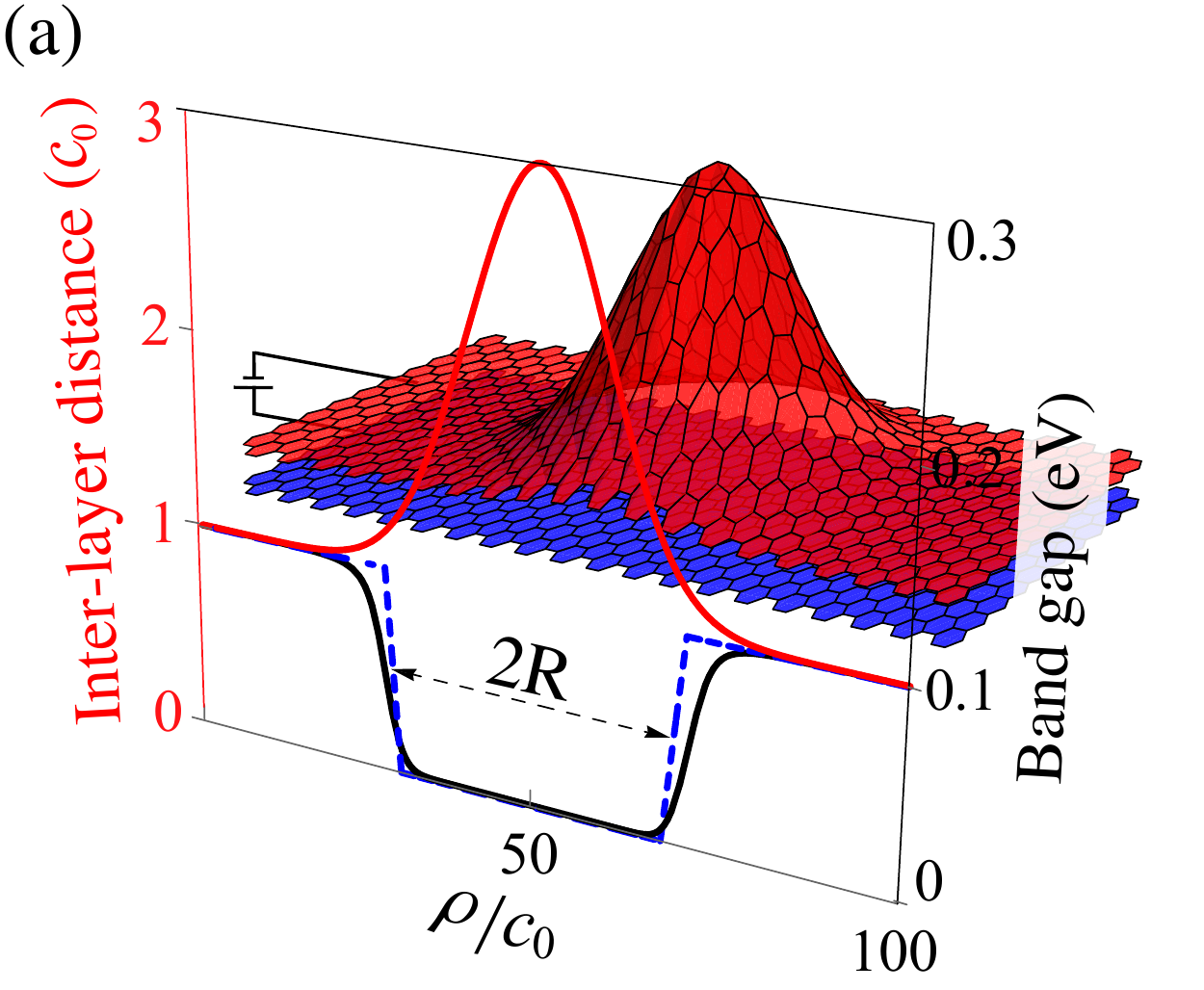}\\
\includegraphics[width=2.8 in]{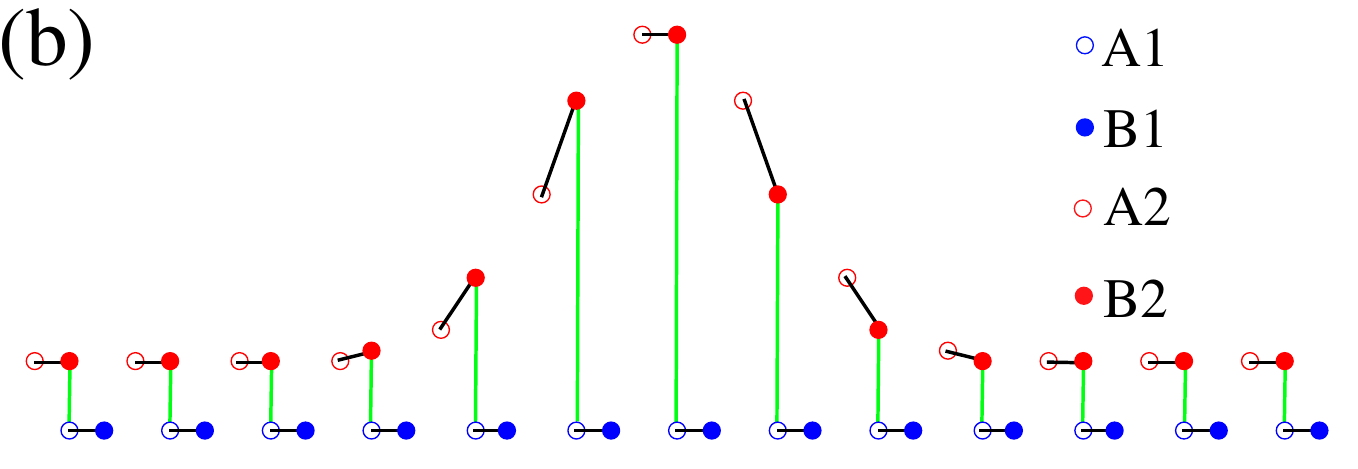}\\
\includegraphics[width=\linewidth]{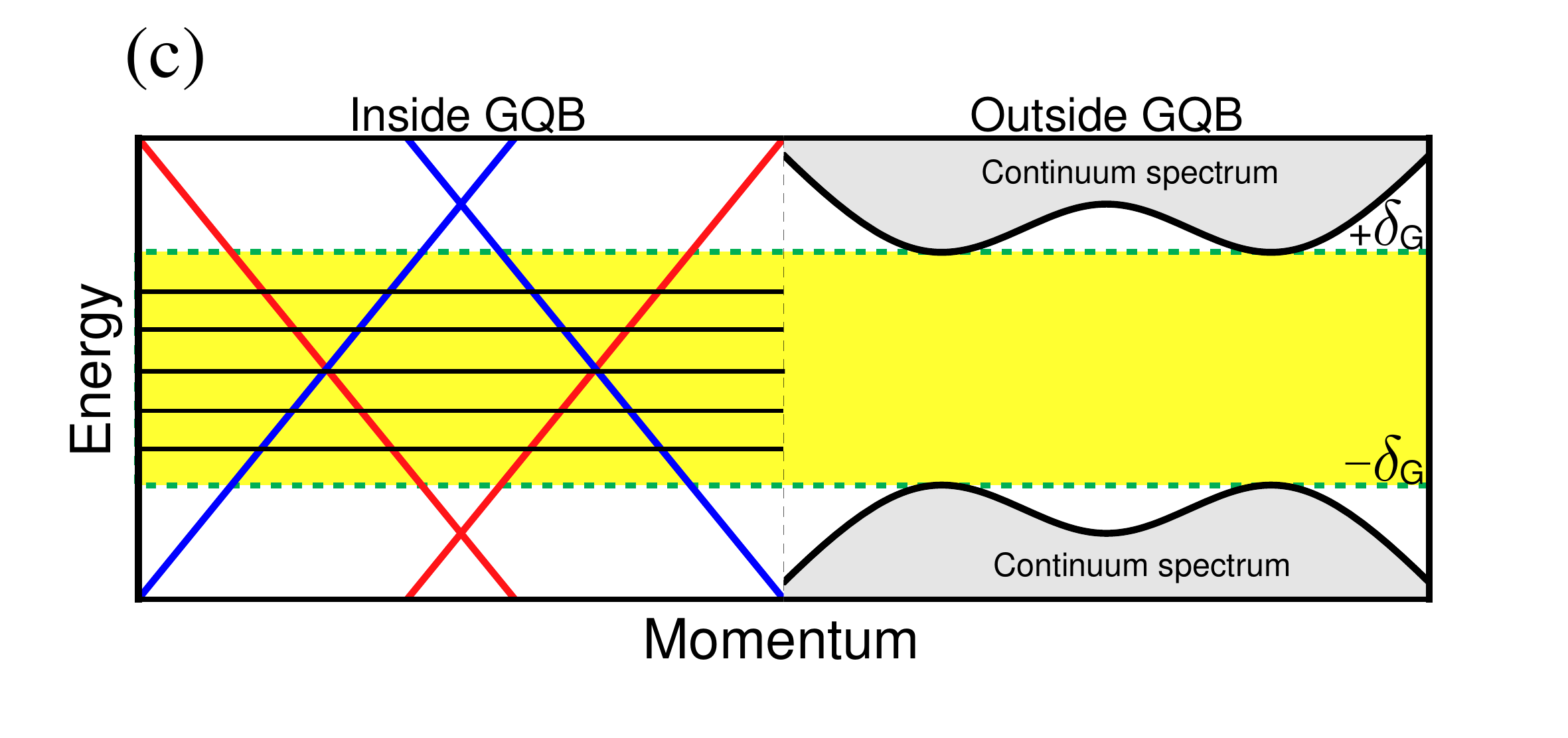}
\caption{(Color online)  (a) Schematic representation of a circular GQB with radius R. The inter-layer distance is shown in red. The black line corresponds to a local band gap for a global bias $\delta=0.12$ eV. The approximate band gap profile with an abrupt change at $\rho=R$ is shown in dashed blue. (b) Schematic representation of a cross section of the GQB depicting the position of the different atoms. The black lines are the $\pi$-orbitals, the vertical green lines represent the inter-layer coupling. For illustrative reasons, only a small number of atoms is shown. The GQBs in this study typically have radii of several hundreds of atoms. (c) Energy spectrum inside (left) and outside (right) the GQB.  Red and blue bands correspond to top and bottom layers while the horizontal black lines in the left figure depict the discrete energy levels occuring due to confinement. These states are only allowed in the range $E<\left\vert \delta_{G} \right\vert$ as delimited by the yellow region in-between the solid black curves that correspond to the edge of the continuum outside the GQB.} \label{fig-GQB}
\end{figure}

Recently, delaminated bilayer graphene (BLG) attracted  attention because of its possibility for layer selective transport \cite{Abdullah2017,Abdullah_2016,Lane2018,Jaskolski2018}. These structures have also been experimentally observed in mechanically exfoliated graphene samples\cite{Yin2016}. GQBs are based on delaminated BLG, but here the delamination is concentrated in a circular region. By application of a global bias gate, states are trapped in this region but retain the interesting graphene-like characteristics. The proposed GQB supports bound states and overcomes the above mentioned limitations. It is free of magnetic fields, relatively possible to manufacture without losing graphene's quality. Finally GQBs also allow external tunability of the electronic spectrum by application of a simple global gate\cite{Ohta_2006} and one can even control the layer localization of the confined states themselves.  

A GQB correstponds to Bernal bilayer graphene where locally  the upper layer is deformed, hence creating a blister in the top layer as shown in Fig. \ref{fig-GQB}. Its electronic spectrum can be probed using STM\cite{Morgenstern2017}, but in contrast to other experiments the electric field of the STM tip is not necessary to confine electrons\cite{Freitag2016,Zhao2015}. As a result of the deformation, the inter-layer coupling strength $\gamma_{1}$ is strongly reduced and practically zero inside the blister. Therefore, the charge carriers have a degenerate linear energy spectrum inside the blister as they belong to independent layers. Outside the blister, however, the two layers are coupled in a Bernal bilayer structure and have the characteristics of a parabolic energy spectrum. By applying a global gate that induces a potential difference between top and bottom layer, a gap can be opened outside the GQB but inside the blister the linear energy spectrum of the two separate layers is only shifted up and down in energy, allowing states for energies in the bilayer gap. These states are bound in the GQB as shown in Fig. \ref{fig-GQB}(c). Since they cannot exist anywhere except in the GQB, the life time of the state diverges and we, therefore, have bound states.

In order to create the blister structure described in the previous paragraph, one could follow several routes. This first one uses the local separation of two graphene layers that is found in several samples \cite{Yan2016,Schmitz2017a,Clark_2014}. By applying a global gate to these of nanostructures, states will confine inside the blister. A second route follows a deliberate introduction atoms in-between two graphene layers with, for example, intercalation techniques\cite{P1988,Kim2011a,Wang2018}. A final route could consist of using graphene samples that are decorated with nanoclusters as a basis material during growth of bilayer graphene. It was shown that current techniques can precisely control over the size and content of these clusters\cite{Scheerder2017}. It is, therefore, expected that by following this technique, we could also precisely control the radius of the GQBs that are made in this way. Creating GQBs as such remains an open quest, but a major advantage of using nanoclusters is that the material type also influences the electronic properties of the confined modes. For example if the nanoclusters are metallic, an dipole will be induced in the nanocluster, which in its turn influences the electric potential felt by the states in a specific and material-dependent way. In this article, to investigate this effect we consider two extreme examples; one in which the nanocluster does not influence the local potential and one in which the local inter-layer potential has been flipped. These results show that whether confinement occurs does not depend on the precise contents of the nanocluster. A detailed modelling of the influence of the nanocluster's material properties on the confined states is, however, beyond the scope of the current article.   

To model the system we use the following Hamiltonian in the continuum limit around the $K$-point \cite{Snyman_2007}:
\begin{equation}\label{starting_Hamiltonian}
\hat{H}(\vec{r})=\left(
\begin{array}{cccc}
    \delta(\vec{r})  & v_{\rm F}\hat{\pi}_{+} &  \gamma_{1}(\vec{r}) & 0 \\
  v_{\rm F}\hat{\pi}_{-} & \delta(\vec{r}) &  0 & 0\\
  \gamma_{1}(\vec{r}) &   0& -  \delta(\vec{r})  & v_{\rm F}\hat{\pi}_{-} \\
  0 & 0& v_{\rm F}\hat{\pi}_{+} & -  \delta(\vec{r})  \\
\end{array}%
\right)~,
\end{equation}
in the basis of orbital eigenstates of the four atoms in the BLG unit cell $\Psi = (\Phi_{A1},\Phi_{B1},\Phi_{B2},\Phi_{A2})^{\dag}$. In Eq. \eqref{starting_Hamiltonian}, $\hat{\pi}_{\pm} = \hat{p}_{x} \pm i \hat{p}_{y}$ is the canonical momentum and the quantity $\delta(\vec{r})$ denotes the potential bias between the two layers induced by the global electrostatic gate. Notice that the latter quantity depends on the position as the structure of the GQB affects the local potential experienced on both layers. $\gamma_{1}(\vec{r})$ is the inter-layer coupling between the $A1$ and $B2$ atoms and, together with $\delta(\vec{r})$, determines the local band gap. A cross section of the atomic configuration is shown in Fig. \ref{fig-GQB}(b). The energy spectrum obtained from Eq. \eqref{starting_Hamiltonian} shows a band gap \cite{Abdullah_2017}
\begin{equation}
\delta_{\rm G}(\vec{r}) = \delta(\vec{r}) \left(1+4\frac{\delta^{2}(\vec{r})}{\gamma_{1}^{2}(\vec{r})}\right)^{-1/2}~.
\label{eq_energy_range}
\end{equation}
The strength of the inter-layer coupling $\gamma_{1}(\vec{r})$ is determined by the distance between both layers. Since the coupling is related to the overlap between the orbital eigenstates of the two carbon atoms right above each other, it decreases exponentially with inter-layer distance. The inter-layer coupling can be written as \cite{Donck2016} %
\begin{equation}
\gamma_{1}(\vec{r}) = \gamma^{0}_{1} \exp\left(-\beta \frac{c(\vec{r})-c_{0}}{c_{0}}\right)~,
\label{gamma1_coupling}
\end{equation}
Here, $\gamma_{1}^{0} = 0.38~{\rm eV}$, $\beta = 13.3$, and $c_{0} \approx 0.3~{\rm nm}$ is the equilibrium inter-layer distance\cite{Lobato_2011, Donck2016}. 

Fig. \ref{fig-GQB}(a) shows the band gap in the GQB as a function of the distance to the center of a gaussian GQB. The result shows that the gap vanishes inside the blister and then increases very sharply at the edge of the GQB. The eigenstates and energy levels of confined states inside the blister can, therefore, be determined by assuming a sharp step in the band gap at $\rho = R$ by matching the different components of the wave functions of the two graphene layers inside with those outside the GQB\cite{Xavier2010}.

\begin{figure}[t!]
\centering\graphicspath{{./Figures/}}
\includegraphics[width=\linewidth]{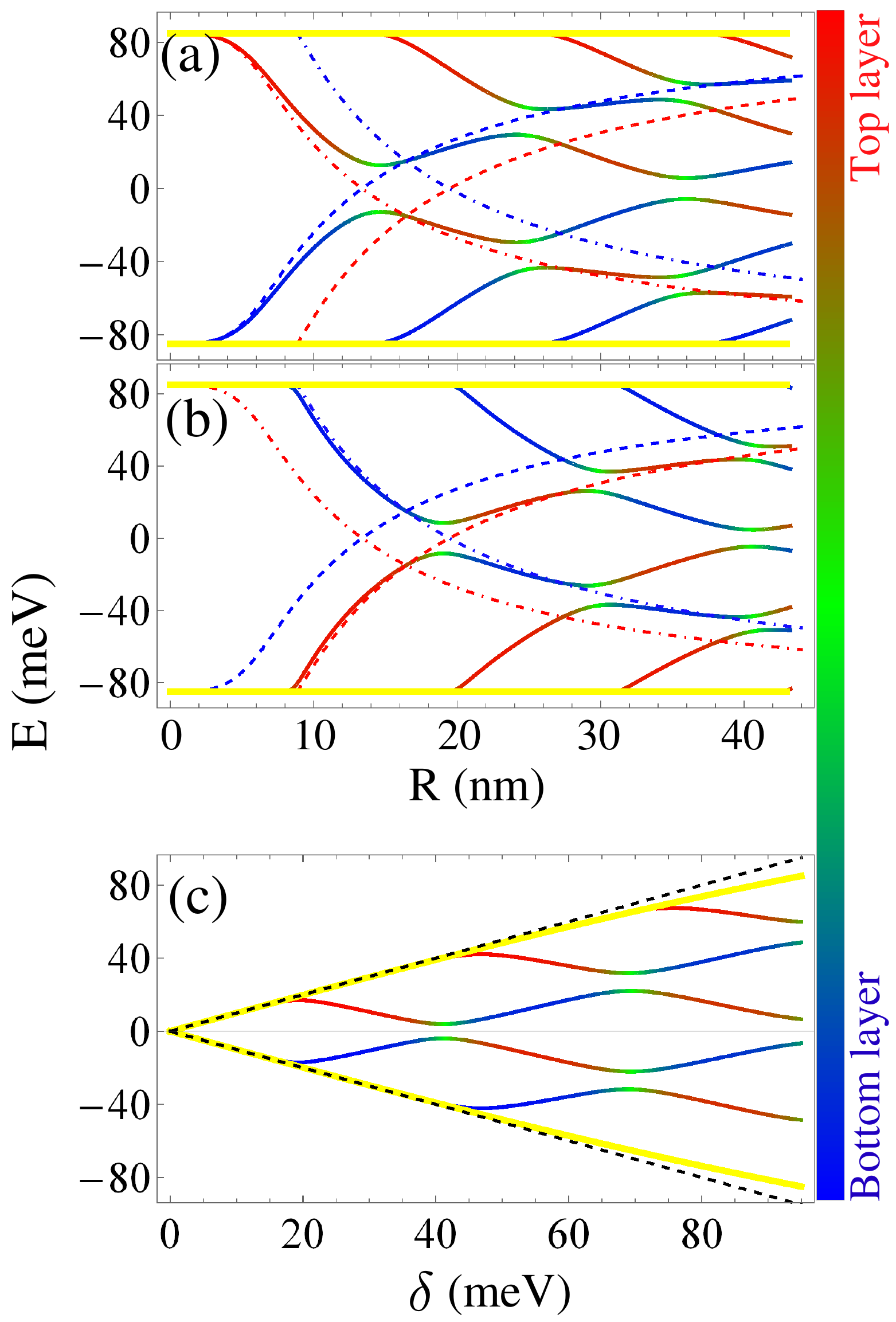}
   \caption{(Color online)  Energy levels of a GQB and corresponding layer occupation indicated by the color for angular quantum number $m=0$. The solid curves in (a) and (b) correspond to a blister with homogenous bias $\delta(\vec{r})$ everywhere or an opposite bias inside the GQB respectively. Yellow horizonal lines delimit the energy range for confinement, i.e. $E=\pm\delta_{G} $. In both graphs $\delta=95$ meV. Dashed and dotted-dashed curves  represent the first energy levels  of  pure  holes and electrons  confined states inside  the blister. (c) Energy levels of GQB  as a function of the global homogeneous bias  for $m=0$ and $R=34.6$ nm. Black dashed and yellow solid curves correspond to $E=\pm \delta$ and $E= \pm \delta_G$, respectively.  } \label{Energy_levels_with_Layer_Occupation}
\end{figure}

The obtained eigenvalues are purely real and thus correspond to  bound states with diverging lifetime. This is a manifestation of the fact that the bias-induced band gap only exists outside the blister. This contrasts with quasi-bound states in the presence of an electrostatic potential in single layer graphene QDs. For these structures, the eigenvalues are complex and thus the states have a finite lifetime\cite{Matulis2008, Hewageegana2008}. 

The resulting energy levels for angular quantum number $m=0$ are shown in Fig. \ref{Energy_levels_with_Layer_Occupation}(a). It consists of discrete energy states that have an oscillatory dependence on the size $R$ of the GQB. For very small radii two in-gap states exist with energy near the edge of the conduction and valence bands. As the size increases, the energy levels approach each other and show avoided crossings. For non-zero angular quantum number similar results are obtained. 

To understand the origin of the energy bands and their anti-crossings, in Fig. \ref{Energy_levels_with_Layer_Occupation}(a) we have color coded the spectrum indicating  the layer to which the corresponding eigenstate belongs. From this it is clear that for small radii the states with positive energy belong to the \textcolor{black}{top} layer, while the negative energy states are positioned at the \textcolor{black}{bottom}. Because the inter-layer bias is applied to the entire sample, also inside the GQB the electronic states are shifted by $-\delta$ or $\delta$ for states on the top or bottom layer, respectively, as shown in the left panel of Fig. \ref{fig-GQB}(c). Therefore, the bottom layer is effectively hole-doped while the top layer is electron-doped due to the bias gate. This is reflected in the behavior of the confined states; indeed the electron state on the top layer decreases in energy as the GQB increases in size, while the hole state at the top layer increases. 

As the two energy levels approach each other, the levels show an anti-crossing at the radius for which the particles are equally distributed over both layers. This happens every time a hole state from the \textcolor{black}{bottom} layer crosses an electron state from the \textcolor{black}{top} layer. The level repulsion is consistent with the Wigner-von Neumann theorem and occurs because the wave functions of both states share the same symmetry  \cite{Greiner1985}. 

As a further proof of the origin of the different energy levels, in Fig. \ref{Energy_levels_with_Layer_Occupation}(a), we also show the energy levels of a pure hole (dashed) and electron (dot-dashed) doped GQB as a function of the  radius of the GQB. This model system is formed by assuming that both layers inside the dot are at the same potential, while outside the blister an inter-layer bias still opens a gap. In Fig. \ref{Energy_levels_with_Layer_Occupation}(a) the energy levels correspond very closely to the numerically calculated results in the GQB. This shows that the anti-crossings result from wave function overlap in the connected bilayer graphene region, i.e. outside the blister. For large GQBs, we find that the energy levels, apart from anti-crossings, can be considered as stemming from two disconnected graphene nanodisks. 

To gain a first insight into how the electrostatic properties of the GQBs nanoclusters could affect the properties of the confined modes, we consider a non-homogenous bias, which is directly visible in the energy levels of the GQB. In Fig. \ref{Energy_levels_with_Layer_Occupation}(b) we show these levels when the bias inside the GQB is exactly opposite to that in the rest of the sample. Since now the top layer is electron doped and the bottom layer is hole doped, the layer occupation is reversed with respect to the previous case with a homogeneous bias. We still observe anti-crossings when electron and hole states become degenerate but they occur at larger radii. Fig. \ref{Energy_levels_with_Layer_Occupation}(b) shows that the electron states now belong to the bottom layer and  the holes belong to the top layer.  Furthermore, the energy levels now correspond to the second branch of the pure electron and hole doped systems as indicated by the dashed and dot-dashed curves.  A detailed description of the effect of specific material's clusters on the confined states is beyond the scope of the present study, but the results shown in Fig. \ref{Energy_levels_with_Layer_Occupation}(b) do show that even for an extreme case of abrupt opposite bias, the confinement is maintained. 

The tunability of GQBs is shown in Fig. \ref{Energy_levels_with_Layer_Occupation}(c) where the energy levels of a GQB of fixed size are shown as a function of inter-layer bias. The result shows that the number of confined energy levels can be tuned over a wide range by simply changing the applied bias gate. These results can be directly verified by local scanning tunnelling microscopy measurements \cite{Lee2016}.

\begin{figure}[t!]
\centering\graphicspath{{./Figures/}}
\includegraphics[width=\linewidth]{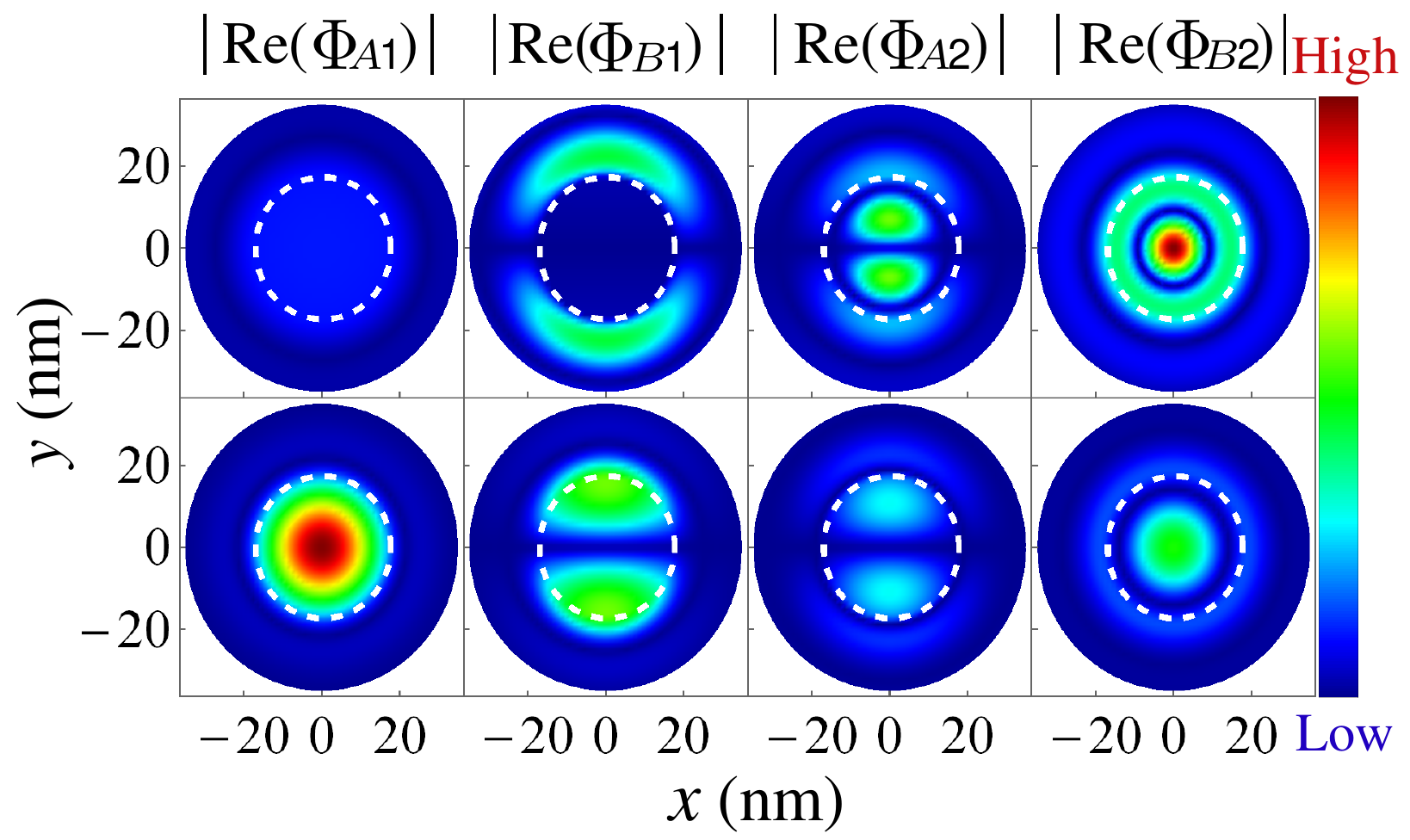}
   \caption{Real part of the different components of the wavefunction for states with energy $E= 76$ meV and $E= 17$ meV for top and bottom rows, respectively. The blister has a radius $R=17.3$ nm   and bias $\delta=95$ meV. The radius of the blister is indicated by a white dashed circle. } \label{Real_Part_Wave}
\end{figure}

Since the inter-layer coupling is active only outside the blister, therefore one expects that the eigenstates are mainly localized inside the radius of the GQB. Peculiarly, however, we find that this is not true for all spinor components of the eigenstate. In Fig. \ref{Real_Part_Wave} we show the real part of the wave function for each component of a blister that supports two positive energy levels. While for the low-energy state (bottom row) the wave function is almost completely localized inside the blister, we see that the high-energy state (top row) has a significant portion outside the blister's radius on the $B1$ component of the bottom layer.

  \begin{figure}[t!]
\centering\graphicspath{{./Figures/}}
\includegraphics[width=\linewidth]{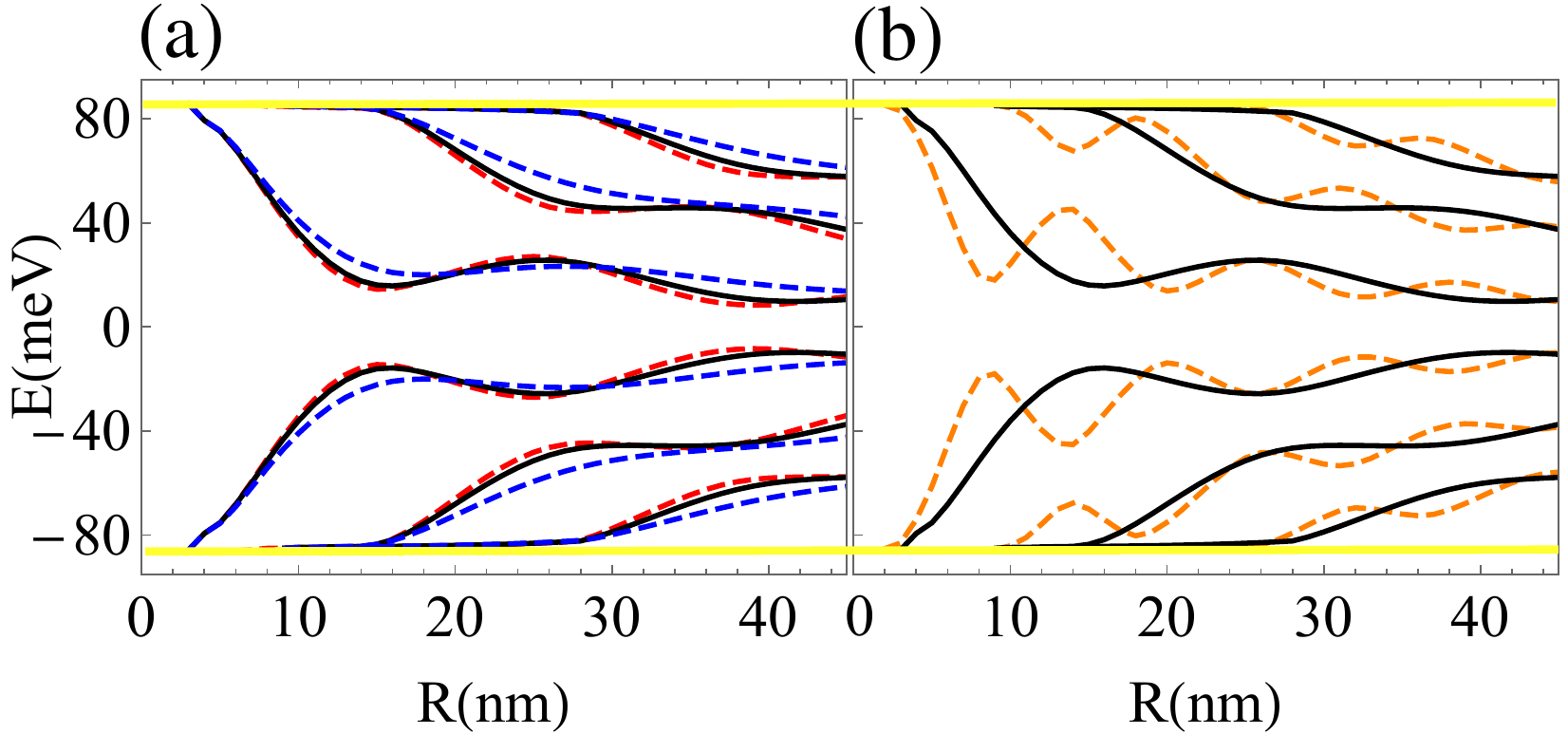}
   \caption{Effect of a gaussian interface to the GQB energy levels. The black curves correspond to the results of Fig.  \ref{Energy_levels_with_Layer_Occupation}(a). (a) Shows a variation of the height of the dome by blue (1.5 $c_0$) and red (10 $c_0$) respectively. In (b) the dashed-orange curve shows the the effect of varying the inter-layer coupling strength through accounting for capacitive effects.  } \label{GQb_Levels_Num}
\end{figure}

Up to now, the analysis was performed by modelling the edge of the blister as an abrupt interface in the band gap describing electronic states. This assumption is justified because of  the very sharp transition between  gapless and gapped states as shown in Fig. \ref{fig-GQB}(a), which is a consequence of the exponential dependence of the inter-layer coupling strength on the inter-layer distance expressed through Eq. \eqref{gamma1_coupling}. Because the band gap changes  sharply at the edge of the system, the morphological details of the blister are obscured and many different shapes of circular blisters effectively have the same energy levels. However, when the height variation of the blister becomes much smaller, this argument might not hold any more. Therefore, in Fig. \ref{GQb_Levels_Num}(a)  we show the energy levels for Gaussian GQBs with varying height. In contrast to the previous analysis for these results we have resorted to numerical calculation of the energy levels using a finite element package. The results show that even small blisters support localized eigenstates with a similar energy spectrum. Notice that the morphology of the blister only affects the strength of the anti-crossings but that already for GQB with a height of twice the equilibrium inter-layer distance, the energy levels are very close to the completely decoupled case discussed above.

We also investigate the effect of a change in inter-layer bias due to capacitive effects. Indeed, since the bias arises due to electrostatic gates, the top layer will be influenced differently when closer to the top gate than the bottom layer. In Fig. \ref{GQb_Levels_Num}(b) we show numerical results (dashed-orange)  for a locally changing inter-layer bias. While also here the confined states result in a robust discretized energy spectrum, the wavelength of the oscillations due to anti-crossings is strongly reduced. This is because in the latter case the cones inside the GQB are shifted more strongly in energy and, therefore, the confined states have a shorter wavelength. 

Finally, note that the deformation of the top layer to form a GQB is in principle associated with a local triaxial strain. Therefore, the inter-atomic distance in the top layer can be slightly larger than the equilibrium distance. This can affect the Fermi velocity $v_{\rm F}$ of the states in the top layer, however, as discussed by Neek-Amal et al\cite{Neek-Amal2013}, triaxial strain will only introduce pseudo-magentic fields near the edge of a finite size graphene flake and in the center it is zero. In our case, the size of the GQB is much smaller than the total size of the bilayer graphene sheet and, therefore, strain has a negligible effect on the results obtained in this study.

In conclusion, graphene quantum blisters are unique electrostatic tunable graphene-based quantum dots. They support bound states with diverging lifetime that can be elegantly realized by  means of only electrostatic gating and are robust against changes in the GQBs morphology. A big advantage of GQBs is the tunability through gate variations. Also, we pointed out that by changing the contents of the blisters, one could access another degree of freedom to establish quantum dot systems with specific energy levels as required for different applications. Therefore, we expect that the GQBs can form the basis of a new subfield in graphene physics where the graphene sheet structure is used together with electric fields to achieve tunable quantum systems.

\section*{Acknowledgments}
HMA and HB acknowledge  the  Saudi
Center for Theoretical Physics (SCTP) for  their generous support and the support of KFUPM under physics research group projects  RG1502-1 and RG1502-2. This work is supported by the Flemish Science Foundation (FWO-Vl) by a post-doctoral fellowship (BVD) and a doctoral fellowship (MVdD).


\end{document}